# Evaluation of the Coulomb and exchange integrals for higher excited states of helium atom, taking into account the interaction between magnetic moments of the electrons.

Voicu Dolocan
*Faculty of Physics, University of Bucharest, Bucharest, Romania*

We have calculated numerically the total energy of the helium atom in higher excited state. The Coulomb and exchange integrals are evaluated via spherical harmonics. The interaction of the magnetic moments of the electrons between them, was taken into account through a *cosine* term in the Coulomb potential. The obtained results are in agreement with experimental data. Lamb shift appears as a natural result of the solution of Schroedinger equation.

## I. INTRODUCTION

Many studies have been devoted to solve Schrodinger equation as accurately as possible for two electron helium atom [1-12]. The perturbation theory, as well as the variational calculation are the most common and valuable methods which achieved a good precision for energy values. Also, the improved hyper-spherical harmonic method and the hyper-spherical coordinate approach have been developed for calculating the energy levels and wave functions directly from the Schrodinger equation in the three-body problem. The helium atom has two electrons and a nucleus of charge +2*e*. The total energy in the case when one of the electrons is in the ground state and the other in a higher excited state characterized by (*nlm*) is

$$E=E_{100}+E_{nlm}+\Delta E \tag{1}$$

where, in the first perturbation theory, $\Delta E$ is obtained by evaluating the expectation value of the potential energy of interaction between the two electrons. It is written

$$\Delta E=C\pm K \tag{2}$$

where $C$ is the Coulomb integral and $K$ is the exchange integral. In the singlet case the spatial function

$$\Psi(r_1,r_2)=\frac{1}{\sqrt{2}}\left[\Psi_{100}(r_1)\Psi_{nlm}(r_2)+\Psi_{100}(r_2)\Psi_{nlm}(r_1)\right] \tag{3}$$

is symmetric, and

$$E_s=E_{100}+E_{nlm}+(C_s+K_s) \tag{4}$$

while, in the triplet case the spatial function

$$\Psi(r_1,r_2)=\frac{1}{\sqrt{2}}\left[\Psi_{100}(r_1)\Psi_{nlm}(r_2)-\Psi_{100}(r_2)\Psi_{nlm}(r_1)\right] \tag{5}$$

is antisymmetric, and

$$E_t = E_{100} + E_{nlm} + \left(C_t - K_t\right) \qquad (6)$$

$\boldsymbol{r}_1$ and $\boldsymbol{r}_2$ are the position vectors of the two electrons, 1 and 2, respectively.

The difference between our paper and the earlier papers is that we add the indices *s* and *t* to the Coulomb and exchange integrals, respectively, because as we will see in the next sections, due to the interaction between spins of the two electrons, we have $C_s < C_t$, $K_s < K_t$, which justifies the Pauli principle. In the next section we will show how is modified the Coulomb's law due to the interaction between magnetic moments of the electrons. Further we will calculate the energy of the higher excited states of helium atom.

## II. SPIN-SPIN INTERACTION AND THE COULOMB'S LAW

The theory and experiments demonstrate that the free electron has a magnetic moment equal to the Bohr magneton $\mu_B$, whose projections on a specified direction are $s_z = \pm\hbar/2 = \hbar m_s$. where $m_s = \pm 1/2$ is the spin quantum number. We write $\mu_z^{(s)} = \mu_B g m_s$ where $g = 2$. We define the vector potential

$$\boldsymbol{A} = \frac{\boldsymbol{\mu}_s \times \boldsymbol{r}}{r^3}$$

where $\boldsymbol{\mu}_s$ is the spin magnetic dipole moment and $\boldsymbol{r}$ is the vector from the electron to an observation point. The magnetic moment generated by the spin is

$$\boldsymbol{\mu}_s = -\frac{eh}{mc}\boldsymbol{m}_s$$

where $\mathbf{m}_s$ is a vector and $m_s$ is the spin quantum number. The wave function for an electron movement along a given path through a magnetic field contains at exponent the term

$$\boldsymbol{q_o}\cdot\boldsymbol{r} - \frac{e}{\hbar c}\oint \boldsymbol{A}.d\cdot\boldsymbol{r}$$

instead of $\mathbf{q}_o.\mathbf{r}$ only. Now, the energy of the electron-electron interaction may be written as[13,14]

$$E_I = -\frac{\hbar^3 D^2}{32 m^2 r^2 \left(\rho_o + \frac{Dr}{c^2}\right)^2} \sum_{q,q_o,k} \frac{\left(\boldsymbol{q}\cdot\boldsymbol{q_o}\right)^2}{\omega_q^2 \omega_{q_o}^2} \frac{1}{2} \left|\sum_n e^{i\left(\boldsymbol{q}_n\cdot\boldsymbol{r}_n - \frac{e}{\hbar}\oint \boldsymbol{A}_n . d\cdot\boldsymbol{r}_n\right)}\right|^2$$

$$\frac{1}{\left(\varepsilon_k - \varepsilon_{k-q}\right) - \omega_q}\left(n_q + 1\right)\left(n_{q_o} + 1\right) n_k n_{k-q} \qquad (7)$$

where *D* is a coupling constant, *m* is the mass of an electron, $\rho_o$ is the massive density of the interacting field, $Dr/c^2$ is the " mass less density" of the interacting field, $\omega_{\mathbf{q}} = cq$ is the classical oscillation frequency of the interacting field, $\omega_{q_o}$ is the oscillation frequency of an electron, $\mathbf{q}$ is the wave vector of the interacting field, $\mathbf{q}_o$ is the wavevector of the boson associated with the electron, **k** is the wave vector of the electron, $\varepsilon_k = \hbar k^2/2\mathrm{m}$ , $n_{qo}$ is the occupation number of the bosons associated with an electron, $n_q$ is the occupation number of the bosons associated with the interacting field, $n_k$ is the occupation number of the electrons. In the case when the interacting field is a photon field, $\rho_o = 0$; for quasi free electron when $\omega_{q_o} = \hbar q_o^2 2\mathrm{m}$ , $\varepsilon_k - \varepsilon_{k-q} << \omega_q$, and for $n_q$, $n_{qo} = 0$, $n_k$, $n_{k-q} = 1$, Eq. (1) becomes

$$E_I = \frac{\hbar^3 c^4}{32 m^2 r^4} \sum_{q,q_o',k} \frac{\left(\boldsymbol{q}\cdot\boldsymbol{q}_o'\right)^2}{\omega_q^2 \omega_{q_o}^2} \frac{1}{2} \left|\sum_n e^{i\cdot \boldsymbol{q}_{on}'\cdot \boldsymbol{r}_n}\right|^2 \tag{8}$$

where

$$\boldsymbol{q}_o' = \boldsymbol{q}_o - \frac{e}{\hbar c}\frac{eh}{mc}\frac{\boldsymbol{m}_s \times \boldsymbol{R}}{R^3} \tag{9}$$

Further,

$$\sum_q \frac{\left(\boldsymbol{q}\cdot\boldsymbol{q}_o'\right)^2}{\omega_q^2 \omega_{q_o'}^2} = \left(\frac{2\mathrm{m}}{\hbar}\right)^2 \frac{1}{q_o 12 c^3} \frac{\Omega}{(2\pi)^2} \int_0^{\pi} \cos^2\alpha \sin\alpha d\alpha \int_0^{q_o'} q dq = \left(\frac{2\mathrm{m}}{\hbar}\right)^2 \frac{r^3}{9\pi c^3}$$

$$\left|\sum_{n=1,2} e^{i\cdot \boldsymbol{q}_n\cdot \boldsymbol{r}_n}\right|^2 = 2\left[1+\cos\Gamma\right]$$

where

$$\Gamma = \boldsymbol{q}_2\cdot\boldsymbol{r}_2 - \boldsymbol{q}_1\cdot\boldsymbol{r}_1 - \frac{e^2}{mc^2}\left(\oint \frac{\boldsymbol{m}_{s2}\times\boldsymbol{r}_2}{r_2^3} d\cdot\boldsymbol{r}_2 - \oint \frac{\boldsymbol{m}_{s1}\times\boldsymbol{r}_1}{r_1^3} d\cdot\boldsymbol{r}_1\right) =$$

$$\frac{1}{2}\left(\boldsymbol{q}_2+\boldsymbol{q}_1\right)\left(\boldsymbol{r}_2-\boldsymbol{r}_1\right) + \frac{1}{2}\left(\boldsymbol{q}_2-\boldsymbol{q}_1\right)\left(\boldsymbol{r}_2+\boldsymbol{r}_1\right) - \Gamma_o$$

$$\Gamma_o = \frac{e^2}{mc^2}\left(\oint \frac{\boldsymbol{m}_{s2}\times\boldsymbol{r}_2}{r_2^3} d\cdot\boldsymbol{r}_2 - \oint \frac{\boldsymbol{m}_{s1}\times\boldsymbol{r}_1}{r_1^3} d\cdot\boldsymbol{r}_1\right) = \frac{e^2}{mc^2}\frac{2\pi}{r}\left(m_{s2}-m_{s1}\right) = \frac{2\pi e^2}{\mathrm{mr}\, c^2}$$

We have considered $m_{s2} = 1/2$, $m_{s1} = -1/2$. For $\mathbf{r}_1 = -\mathbf{r}_2$ and $\boldsymbol{r}_1 = \boldsymbol{r}_2 = \boldsymbol{r}$, $\mathbf{q}_1 = \mathbf{q}_2 = \mathbf{q}_o$ we write

$$\Gamma_o = q_o r\cos\theta - \Gamma_o = q_o' r\cos\theta - \Gamma_o \tag{10}$$

By using the integral

$$\int_0^{\pi} \cos\left(q_o' r\cos\theta - \Gamma_o\right)\sin\theta d\theta = 2\cos\left(\Gamma_o\right)\frac{\sin\left(q_o' r\right)}{q_o' r}$$

and integrating over $q'_o$ from 0 to $0.94\pi/$ R, one obtains

$$E_I=\frac{\hbar c}{144\pi r}\left[2+1.3\cos\left(\frac{2\pi e^2}{mrc^2}\right)\right] \qquad (11)$$

For $\Gamma_o = 0$, one obtains just Coulomb's law

$$E_I=\frac{\alpha\hbar c}{r} \qquad (12)$$

where $\alpha$ = 1/137 is the fine structure constant. The upper limit of the integral over $q'_o$ was chosen for the agreement of value of $\alpha$ with experimental data. We specify that the upper limit of $q'_o$ should be of 0.76π/r , as it is imposed by the constraint condition $\left[\left(4\pi r^3/3\right)/\left(2\pi\right)^3\right]\times 4\pi q_o^3/3=1$ . This given that $\alpha$ = 1/147, which differ some few from the fine structure constant. It is observed that when the spins of the two interacting electrons are anti-parallel, the interacting energy is modulated by a term *cosine* which depends on the spin magnetic moment and on the distance between the two electrons.

The energy levels of atomic electrons are affected by the interaction between the electron spin magnetic moment and the orbital magnetic moment of the electron. It can be visualized as a magnetic field caused by the electron's orbital motion interacting with the spin magnetic moment. The effective magnetic field can be expressed in terms of the electron orbital angular momentum. We consider the vector potential $\boldsymbol{A}=\frac{\boldsymbol{\mu}\times\boldsymbol{r}}{r^3}$ where $\boldsymbol{\mu}$ is the magnetic dipole moment and $\boldsymbol{r}$ is a vector from the middle of the loop to an observation point. An electron in a stationary state in an atom, having a definite angular momentum projection $L_z = \hbar m_l$ ($m_l$ the quantum magnetic number), has a magnetic moment $\mu_z^{(l)}=\mu_B m_l$ where $\mu_B=e\hbar/2mc$ is the Bohr magneton. In these conditions $\Gamma_o$ becomes

$$\Gamma_o=\frac{e^2}{2mc^2}\left[\frac{2\pi}{r_2}m_{l2}-\frac{2\pi}{r_1}m_{l1}\right]+\frac{2\pi e^2}{mr_{12}c^2}\left[m_{s2}-m_{s1}\right]$$

In this case $r_1$ and $r_2$ are the radii of the electrons 1 and 2, respectively and $r_{12}=\left|r_1-r_2\right|$ .

## III. EVALUATION OF THE COULOMB AND EXCHANGE INTEGRALS FOR HIGHER EXCITED STATES OF HELIUM ATOM BY TAKING INTO ACCOUNT THE INTERACTION BETWEEN MAGNETIC MOMENTS OF THE ELECTRONS.

We calculate the Coulomb integral and the exchange integral for higher excited states of helium atom by using spherical harmonics approach. The Coulomb integral can be written as

$$C=\iint\left|\Psi_{100}\left(r_1\right)\right|^2\frac{\hbar c}{144\pi\left|r_2-r_1\right|}\left[2+1.3\cos\left(\Gamma_o\right)\right]\left|\Psi_{nlm}\left(r_2\right)\right|^2 d^3r_1 d^3r_2 \qquad (14)$$

where $\Gamma_o$ is given by expression (13). The exchange integral can be written as

$$K=\iint \Psi^{+}_{100}(r_2)\Psi^{+}_{nlm}(r_1)\frac{\hbar c}{144\pi|r_2-r_1|}\left[2+1.3\cos(\Gamma_o)\right]\Psi_{100}(r_1)\Psi_{nlm}(r_2)d^3r_1d^3r_2 \tag{15}$$

where

$$\Psi_{nlm}(r)=R_{nl}(r)Y_{lm}(\theta,\varphi)$$
$$d^3r=r^2drd\Omega=r^2dr\sin\theta d\theta d\varphi$$

The wave functions used in our calculations are given in Appendix. The obtained results are presented in Table I. The values of $\Gamma_o$ (13) *cosine* argument are given in Table II. We have used non dimensional quantities $x_i = r_i/a_o$, where $a_o$ is Bohr radius. The obtained results are in agreement with experimental data[11,12, 15,16], which are also written in Table I. The integrals were performed in Mathematica 5.2. We believe that by using a better accuracy ofcalculation, a splitting of the the $m_l$ levels one obtains, in according to the values of $\Gamma_o$ (Table II).

## IV. CONCLUSIONS

We have calculated the total energy of the helium atom in higher excited state . The Coulomb and exchange integrals were evaluated via spherical harmonics. We have introduced the interaction between the magnetic moments of the electrons via a term *cosine* in the potential of interaction. The obtained values of the energy levels are in agreement with experimental data. The Lamb shift appears as a natural result of the solution of Schroedinger equation. In another paper we have calculated the energy levels of the states in hydrogen atom, and likewise we have obtained the Lamb shift as a natural result of the solution of Schroedinger equation with modified Coulomb potential[17].

## APPENDIX

The radial wave functions $R_{nl}$ used in our calculations are presented in Table III. Further,

$$Y_{oo}(\theta_1,\varphi_1)=\frac{1}{\sqrt{4\pi}}$$
$$\int |Y_{lm}(\theta_2,\varphi_2)|^2\sin\theta_2 d\theta_2 d\varphi_2=1$$

Table I. The Coulomb integral C, the exchange integral K and the total energy E of the helium atom in higher excited states

| Spectral term | *C*,eV | *K*,eV | *E*,eV | *Experimental* |
|---|---|---|---|---|
| 1s2s $^1$S | 11.41573333967 | 1.19033533943 | -57.67460200476 | -58.3624572424563 |
| 1s3s $^1$S | 5.411489756968 | 0.313731985586 | -54.63033380901 | -56.0579202431514 |
| 1s4s $^1$S | 3.148677499813 | 0.127820901469 | -54.43850159872 | -55.2491296415375 |
| 1s5s $^1$S | 2.040721787947 | 0.066826210720 | -54.38525200133 | |
| 1s6s $^1$S | 1.433422772975 | 0.03294316262 | -56.35876043217 | |
| 1s2s $^3$S | 11.41569905002 | 1.193796374161 | -55.29050457582 | -59.1581011131610 |
| 1s3s $^3$S | 5.411510073834 | 0.313734833949 | -5525778031586 | -56.2604609645120 |
| 1s4s $^3$S | 3.148601997296 | 0.127821649383 | -54.69421966209 | -55.3634244356413 |
| 1s5s $^3$S | 2.040718737378 | 0.068655072945 | -54.52073633557 | |
| 1s6s $^3$S | 1.433427673648 | 0.036705879762 | -55.43216709499 | |
| 1s2p $^1$P | 13.20668283822 | 4.377358489379 | -50.3159586724 | -57.7604441467894 |
| 1s3p $^1$P | 5.919429448189 | 1.132920895331 | -53.30320529204 | -55.8915139004001 |
| 1s4p $^1$P | 3.34175054095 | 0.50625293162 | -53.86699662743 | -55.192768960000 |
| 1s5p $^1$P | 2.13804836033 | 0.212860084362 | -54.14189155531 | |
| 1s6p $^1$P | 1.573228618317 | 0.131393496193 | -54.12476679210 | |
| 1s2p $^3$P | 13.20668263827 | 4.377369917556 | -59.07068707934 | -58.0143534226402 |
| 1s3p $^3$P | 5.919445500939 | 1.132928039069 | -55.56903809369 | -55.9717377584023 |
| 1s4p $^3$P | 3.346098605808 | 0.506255610099 | -54.87515700439 | -55.2361266080000 |
| 1s5p $^3$P | 2.14811373709 | 0.212860604472 | -54.55754686738 | |
| 1s6p $^3$P | 1.494869376612 | 0.131393962756 | -54.46541347503 | |
| 1s3d $^1$D | 6.038936224483 | 0.138938872622 | -54.17768040845 | -55.90432560000 |
| 1s4d $^1$D | 3.394491631363 | 0.073433630202 | -54.24707473844 | -55.185460112000 |
| 1s5d $^1$P | 2.179248482903 | 0.040714971468 | -54.27883654562 | |
| 1s6d $^1$D | 1.5578325773792 | 0.024486715885 | -54.24656959562 | |
| 1s3d $^3$D | 6.038954110872 | 0.13893929351 | -54.45554073819 | -55.904782880000 |
| 1s4d $^3$D | 3.396896310627 | 0.073433808768 | -54.39151743814 | -55.187986721000 |
| 1s5d $^3$D | 2.174201938516 | 0.040715081461 | -54.36531314295 | |
| 1s6d $^3$D | 1.509987410785 | 0.024480785005 | -54.34338826311 | |
| 1s4f $^1$F | 3.399099891361 | 0.009643500679 | -54.30625660796 | |
| 1s5f $^1$F | 2.185613795383 | 0.001574564257 | -54.3116164036 | |
| 1s6f $^1$F | 1.605041072059 | 0.001281489028 | -54.2225663269 | |
| 1s4f $^3$F | 3.399576055473 | 0.00964750493 | -54.32506744946 | |
| 1s5f $^3$F | 2.175724708252 | 0.001574564757 | -54.32464985501 | |
| 1s6f $^3$F | 1.6076422135588 | 0.001281492409 | -54.22252824371 | |
| 1s5g $^1$G | 2.178870296447 | 0.000015726533 | -54.31391397702 | |
| 1s6g $^1$G | 1.483713154824 | 0.000016880836 | -54.34515885323 | |
| 1s5g $^3$G | 2.175749816148 | 0.000015726658 | -54.316466591051 | |
| 1s6g $^3$G | 1.510933520492 | 0.000016880972 | -54.31797224917 | |
| 1s6h $^1$H | 0.051672384008 | 0.000000002782 | -55.7772165021 | |
| 1s6h $^3$H | 0.051938436641 | 0.000000002782 | -55.77695345509 | |

Table II. The values of *cosine* argument $\Gamma_o$; $S = m_{s1} + m_{s2}$

| S | $m_l$ | $\Gamma_o$ |
|---|---|---|
| 0 | 0 | $\frac{3.3839\times10^{-4}}{\sqrt{x_1^2+x_2^2-2x_1x_2\cos\theta_1}}=\Gamma_1$ |
| 0 | -1 | $\Gamma_1+\frac{1.6919\times10^{-4}}{x_2}$ |
| 0 | -2 | $\Gamma_1+\frac{3.3839\times10^{-4}}{x_2}$ |
| 0 | -3 | $\Gamma_1+\frac{5.0758\times10^{-4}}{x_2}$ |
| 0 | -4 | $\Gamma_1+\frac{6.6768\times10^{-4}}{x_2}$ |
| 0 | -5 | $\Gamma_1+\frac{8.4597\times10^{-4}}{x_2}$ |
| 0 | 5 | $\Gamma_1-\frac{8.4597\times10^{-4}}{x_2}$ |
| 0 | 4 | $\Gamma_1-\frac{6.6768\times10^{-4}}{x_2}$ |
| 0 | 3 | $\Gamma_1-\frac{5.0758\times10^{-4}}{x_2}$ |
| 0 | 2 | $\Gamma_1-\frac{3.3839\times10^{-4}}{x_2}$ |
| 0 | 1 | $\Gamma_1-\frac{1.6919\times10^{-4}}{x_2}$ |
| 1 | 0 | 0 |
| 1 | -1 | $-\frac{1.6919\times10^{-4}}{x_2}$ |
| 1 | -2 | $-\frac{3.3839\times10^{-4}}{x_2}$ |
| 1 | -3 | $-\frac{5.0758\times10^{-4}}{x_2}$ |
| 1 | -4 | $-\frac{6.6768\times10^{-4}}{x_2}$ |
| 1 | -5 | $-\frac{8.4597\times10^{-4}}{x_2}$ |
| 1 | 5 | $\frac{8.4597\times10^{-4}}{x_2}$ |
| 1 | 4 | $\frac{6.6768\times10^{-4}}{x_2}$ |
| 1 | 3 | $\frac{5.0758\times10^{-4}}{x_2}$ |
| 1 | 2 | $\frac{3.3839\times10^{-4}}{x_2}$ |
| 1 | 1 | $\frac{1.6919\times10^{-4}}{x_2}$ |

Table III. Radial wave functions $R_{nl}(r)$

| *n* | *l* | $R_{nl}(l)$ |
|---|---|---|
| 1 | 0 | $2\left(\frac{2}{a_o}\right)^{3/2} e^{-2r/a_o}$ |
| 2 | 0 | $\left(\frac{1}{a_o}\right)^{3/2}\left(2-\frac{2r}{a_o}\right)e^{-r/a_o}$ |
| 2 | 1 | $\left(\frac{1}{a_o}\right)^{3/2}\frac{2}{\sqrt{3}}\frac{r}{a_o}e^{-r/a_o}$ |
| 3 | 0 | $2\left(\frac{2}{3a_o}\right)^{3/2}\left[1-\frac{4}{3}\frac{r}{a_o}+\frac{8}{27}\left(\frac{r}{a_o}\right)^2\right]e^{-2r/3a_o}$ |
| 3 | 1 | $\left(\frac{2}{3a_o}\right)^{3/2}\frac{4\sqrt{2}}{3}\left(1-\frac{2}{3}\frac{r}{a_o}\right)\frac{2r}{3a_o}e^{-2r/3a_o}$ |
| 3 | 2 | $\left(\frac{2}{3a_o}\right)^{3/2}\frac{2\sqrt{2}}{27\sqrt{5}}\left(\frac{2r}{a_o}\right)^2 e^{-2r/3a_o}$ |
| 4 | 0 | $\frac{1}{96}\left(\frac{2}{a_o}\right)^{3/2}\left[24-36\frac{r}{a_o}+12\left(\frac{r}{a_o}\right)^2-\left(\frac{r}{a_o}\right)^3\right]e^{-r/2a_o}$ |
| 4 | 1 | $\frac{1}{32\sqrt{15}}\left(\frac{2}{a_o}\right)^{3/2}\left[20-10\frac{r}{a_o}+\left(\frac{r}{a_o}\right)^2\right]\frac{r}{a_o}e^{-r/2a_o}$ |
| 4 | 2 | $\frac{1}{96\sqrt{5}}\left(\frac{2}{a_o}\right)^{3/2}\left(6-\frac{r}{a_o}\right)\left(\frac{r}{a_o}\right)^2e^{-r/2a_o}$ |
| 4 | 3 | $\frac{1}{96\sqrt{35}}\left(\frac{2}{a_o}\right)^{3/2}\left(\frac{r}{a_o}\right)^3e^{-r/2a_o}$ |
| 5 | 0 | $\frac{1}{300\sqrt{5}}\left(\frac{2}{a_o}\right)^{3/2}\left[120-192\frac{r}{a_o}+76.8\left(\frac{r}{a_o}\right)^2-10.24\left(\frac{r}{a_o}\right)^3+0.4096\left(\frac{r}{a_o}\right)^4\right]e^{-2r/5a_o}$ |
| 5 | 1 | $\frac{1}{150\sqrt{30}}\left(\frac{2}{a_o}\right)^{3/2}\left[120-72\frac{r}{a_o}+\frac{288}{125}\left(\frac{r}{a_o}\right)^2-\frac{64}{125}\left(\frac{r}{a_o}\right)^3\right]\frac{4}{5}\frac{r}{a_o}e^{-2r/5a_o}$ |
| 5 | 2 | $\frac{1}{150\sqrt{70}}\left(\frac{2}{a_o}\right)^{3/2}\left[42-\frac{56}{5}\frac{r}{a_o}+\frac{16}{25}\left(\frac{r}{a_o}\right)^2\right]\frac{16}{25}\left(\frac{r}{a_o}\right)^2e^{-2r/5a_o}$ |
| 5 | 3 | $\frac{1}{300\sqrt{70}}\left(\frac{2}{a_o}\right)^{3/2}\left(8-\frac{4}{5}\frac{r}{a_o}\frac{64}{125}\right)\left(\frac{r}{a_o}\right)^3e^{-2r/5a_o}$ |
| 5 | 4 | $\frac{1}{900\sqrt{70}}\left(\frac{2}{a_o}\right)^{3/2}\frac{256}{625}\left(\frac{r}{a_o}\right)^4e^{-2r/5a_o}$ |
| 6 | 0 | $\frac{1}{2160\sqrt{6}}\left(\frac{2}{a_o}\right)^{3/2}[720-1200\frac{r}{a_o}+\frac{1600}{3}\left(\frac{r}{a_o}\right)^2-\frac{800}{9}\left(\frac{r}{a_o}\right)^3+$ $\frac{160}{27}\left(\frac{r}{a_o}\right)^4-\frac{32}{243}\left(\frac{r}{a_o}\right)^5]e^{-r/3a_o}$ |
| 6 | 1 | $\frac{1}{432\sqrt{210}}\left(\frac{2}{a_o}\right)^{3/2}\left[840-560\frac{r}{a_o}+112\left(\frac{r}{a_o}\right)^2-\frac{224}{27}\left(\frac{r}{a_o}\right)^3+\frac{16}{81}\left(\frac{r}{a_o}\right)^4\right]\times\frac{2}{3}\frac{r}{a_o}e^{-r/3a_o}$ |
| 6 | 2 | $\frac{1}{864\sqrt{105}}\left(\frac{2}{a_o}\right)^{3/2}\left[336-112\frac{r}{a_o}+\frac{32}{3}\left(\frac{r}{a_o}\right)^2-\frac{8}{27}\left(\frac{r}{a_o}\right)^3\right]\frac{4}{9}\left(\frac{r}{a_o}\right)^2e^{-r/3a_o}$ |
| 6 | 3 | $\frac{1}{2592\sqrt{35}}\left(\frac{2}{a_o}\right)^{3/2}\left[72-12\frac{r}{a_o}+\frac{4}{9}\left(\frac{r}{a_o}\right)^2\right]\frac{8}{27}\left(\frac{r}{a_o}\right)^3e^{-r/3a_o}$ |
| 6 | 4 | $\frac{1}{12960\sqrt{7}}\left(\frac{2}{a_o}\right)^{3/2}\left(10-\frac{2}{3}\frac{r}{a_o}\right)\frac{16}{81}\left(\frac{r}{a_o}\right)^{4}e^{-r/3a_o}$ |
| 6 | 5 | $\frac{1}{12960\sqrt{77}}\left(\frac{2}{a_o}\right)^{3/2}\frac{32}{243}\left(\frac{r}{a_o}\right)^{5}e^{-r/3a_o}$ |